\documentclass[fleqn,usenatbib]{mnras}

\usepackage{newtxtext,newtxmath}

\usepackage[T1]{fontenc}
\usepackage{ae,aecompl}
\usepackage{times}

\usepackage{graphicx}	
\usepackage{amsmath}	
\usepackage{amssymb}	

\newcommand{\rd}{\hspace{1pt}\textrm{d}}

\newcommand{\erf}[1]{\text{erf}\left( {#1} \right)}

\newcommand{\vc}[1]{\mathbf{#1}}
\newcommand{\change}[1]{\textcolor{black}{#1}}

\title[Models of Exponential Bars]{Models of Bars II: Exponential Profiles}

\author[McGough et al]{D.~P.~McGough,$^{1}$\thanks{E-mail: dpm40@alumni.cam.ac.uk, nwe@ast.cam.ac.uk, jls@ast.cam.ac.uk}
N.~W.~Evans,$^{1}$
J.~L.~Sanders$^{1}$
\\
$^{1}$Institute of Astronomy, Madingley Rd, Cambridge CB3 0HH\\
}

\date{Accepted XXX. Received YYY; in original form ZZZ}

\pubyear{2015}

\begin{document}
\label{firstpage}
\pagerange{\pageref{firstpage}--\pageref{lastpage}}
\maketitle

\begin{abstract}
	We present a new model for galactic bars with exponentially falling major axis luminosity profiles and Gaussian cross-sections. This is based on the linear superposition of Gaussian potential-density pairs with an exponential weight function, using an extension of the method originally introduced by \cite{Lo92}. We compute the density, potential and forces, using Gaussian quadrature. These quantities are given as explicit functions of position. There are three independent scaled bar parameters that can be varied continuously to produce bespoke bars of a given mass and shape. We categorise the effective potential by splitting a reduced parameter space into six regions. Unusually, we find bars with three stable Lagrange points on the major axis are possible. Our model reveals a variety of unexpected orbital structure, including a bifurcating $x_1$ orbit coexisting with a stable $x_4$ orbit. Propeller orbits are found to play a dominant role in the orbital structure, and we find striking similarities between our bar configuration and the model of \cite{Ka96}. We find a candidate orbital family, sired from the propeller orbits, that may be responsible for the observed high velocity peaks in the Milky Way's bar. As a cross-check, we inspect, for the first time, the proper motions of stars in the high velocity peaks, which also match our suggested orbital family well. This work adds to the increasing body of evidence that real galactic bars may be supported at least partly by propeller orbits rather than solely the $x_1$ family.
\end{abstract}




\section{Introduction}

A central bar-shaped concentration of stars is a common feature of spiral galaxies. There is ample evidence to suggest that the Milky Way is a late-type barred spiral galaxy from infrared photometry, starcounts, stellar and gas kinematics and microlensing \citep[e.g.,][]{We92,Pa94,Dw95,We15,Sa19a}. This is no surprise, as it has been known since the 1960s that bars form readily in $N$-body simulations, and are long-lived, robust stellar dynamical equilibria~\citep{Ho71,To81}. 

Analyses of observational evidence from photometry of barred galaxies suggests there are in fact two main types of galactic bars - ``flat" and ``exponential" \citep{El85,Se93}. In early-type barred galaxies, the luminosity along the major axis falls slowly, and is sometimes almost flat all the way to the end of the bar. In contrast, late-type barred galaxies have bars with an almost exponentially falling luminosity profile along the major axis. This dichotomy may even be an evolutionary sequence, as early-type barred galaxies tend to be more massive. In this picture, exponential bars may gradually redistribute their mass and angular momentum to become flatter in profile. \change{By contrast, there have been few systematic studies of the density profiles along the minor and intermediate axes, which of course requires deprojection of the surface photometry. The sparse information that we possess suggests that their profiles appear to be close to Gaussian~\citep{Bl83} or exponential~\citep{Ga07}.}

As measured against the endpoints of $N$-body experiments, the analytic models used to describe bars are often unrealistic. \change{For example, the analytic \citet{Fe77} ellipsoids have zero density outside a given elliptical radius~\citep{BT}, namely
\begin{equation}
\rho(x,y,z) = \begin{cases}\rho_0 (1-m^2)^n & m<1 \\
                           0 & m\geq 1, \end{cases}
\label{eq:ferrers}
\end{equation}
with $m^2 = x^2/a^2 + y^2/b^2 + z^2/c^2$. Here $a,b$ and $c$ are the constant semiaxes of the ellisoidal density contours, whilst $n$ is an integer. The gravitational potential within the bar is a polynomial of order $2n+2$ in $x,y$ and $z$. Their tractability means there have been extensive investigations of the orbital structure of Ferrers bars~\citep[e.g.,][]{Pf84a, At92, Sk02}. On the other end of the spectrum, the purely numerical Cazes bar \citep{Ca00,Ba01} is constructed from realistic hydrodynamical simulations, but has a potential defined only on an 800$\times$800 Cartesian grid, and so is not simple to investigate. A huge amount of insight into orbital properties of bars has been discovered through investigation of these models, and their usefulness should not be underestimated. Despite this, the inventory of realistic and simply calculable bar models is small, and a gap in the market remains. This is especially the case for exponential bars, for which there are no simple models in the literature. For example, although Ferrers bars with large values of $n$ have more rapid density fall-off, it is always polynomial, and never exponential.}

\cite{Lo92} introduced a convenient and versatile algorithm for producing flexible barred potentials. They convolved a simple spherical or axisymmetric background potential with a needle-like weight function. They applied their method to the Plummer and Miyamoto-Nagai models to produce a variety of prolate and triaxial bars. Their method was then used successfully in \citet{Wi17} to produce a model for a flat bar by convolving a logarithmic density with a needle-like weight function. Here, we will extend the \citet{Lo92} algorithm, but with an exponential weight function with a view to producing exponential bars. 

The paper is arranged as follows. Section~\ref{section:building_the_bar} presents some simple properties of our new model and calibrates it against an $N$-body simulation that mimics the Milky Way bar. The orbital structure is discussed in some detail in Section~\ref{section:eff_pot_structure} via characteristic diagrams and Poincar\'e surfaces. Section~\ref{section:hvpeaks} presents an application to the presence of high velocity peaks in the Milky Way bulge. \change{We sum up in Section~\ref{section:conclusion}, outlining some future challenges.}

\begin{figure*}
	\centering
	\includegraphics[height=5cm]{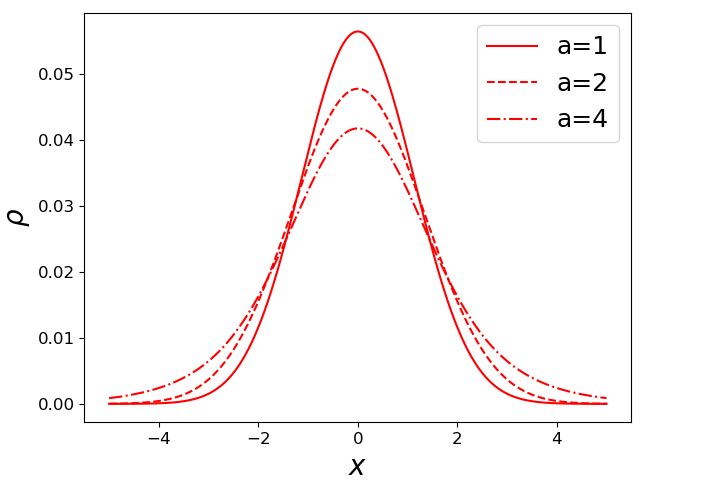} 
    \includegraphics[height=5cm]{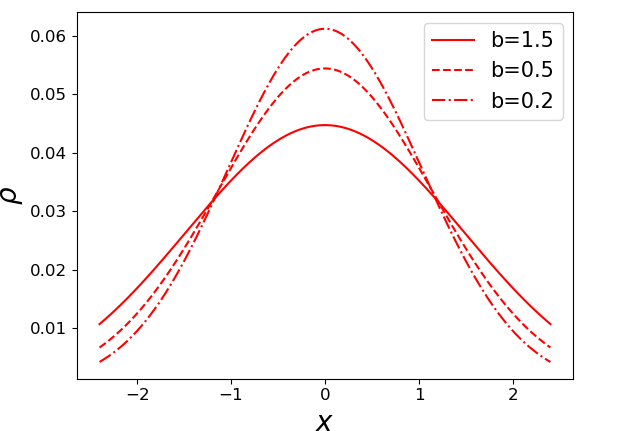} 
	\caption{Left: Major axis density profiles, showing that as $a$ increases, the length of the bar increases, but the central peaked shape remains present. The model has ($b, \sigma,\epsilon) = (1,1,0)$. Right: Major axis density profiles, showing that as $b$ increases, the bar's density profile becomes flatter. The model has ($a,\sigma,\epsilon) = (2,1,0)$.}
	\label{fig:rho_x_profile_2}
\end{figure*}

\section{Model Building}\label{section:building_the_bar}

\cite{Lo92} introduced a method for constructing barred potential-density pairs from a general axisymmetric potential-density pair $(\Phi_a,\rho_a)$ via convolution with a weight function $w(x)$:
\begin{equation}
\begin{split}
\rho_{\text{bar}} = \int w(x')\rho_{a}(x-x',y,z)\rd x',\\
\Phi_{\text{bar}} = \int w(x')\Phi_{a}(x-x',y,z)\rd x' .
\end{split}
\end{equation}
The weight function is a one-dimensional function of the major axis coordinate, such that the new density is preferentially stretched along the $x$ direction.

Our aim is to build a triaxial bar with a major axis density profile that is roughly exponential, and then derive the potential and the forces acting on a particle due to the density distribution. \change{Taking the minor and intermediate axis profiles as Gaussian in cross-section, we start with the density ansatz}
\begin{gather}
\rho_{\text{Gaussian}} = \frac{M}{q\left(2\pi\sigma^2\right)^{3/2}}\exp\left[-\left(x^2 + y^2 + z^2/q^2\right)/\left(2\sigma^2\right)\right] 
\end{gather}
with total mass $M$, variance $\sigma^2$ and flattening ratio $q$. The gravitational potential of the Gaussian is \citep{Ca08}
\begin{gather}
\Phi_{\text{Gaussian}} = -\sqrt{\frac{2}{\pi\sigma^2}}GM\int_{0}^{1}H(m)\rd m , \label{eq:phi_g}
\end{gather}
where
\begin{gather}
H(m) = \frac{1}{\sqrt{1-\epsilon^2m^2}}\exp\left[-\frac{m^2}{2\sigma^2}\left(R^2 + \frac{z^2}{1-\epsilon^2m^2}\right)\right] , \label{eq:h_m}
\end{gather}
with $\epsilon$ given by $\epsilon^2 = 1-q^2$. Here, $G$ is the gravitational constant. This result follows from the general potential theory of ellipsoidal density distributions \citep{EFE}. 
In constructing their bar, \cite{Wi17} used constant weight functions ($w(x)=1$ for $-a<x<a$), producing flattish major-axis density profiles, suitable for early-type bars. Here, we instead use an exponential weight function, to produce near-exponential profiles suitable for late-type bars. We use the weight function
\begin{gather}
w(x) = \frac{1}{2b(1-e^{-a/b})}\exp\left[-\frac{|x|}{b}\right]
\end{gather}
in the region $-a<x<a$, where our pre-factor ensures that 
\begin{gather}
\int_{-a}^{a}w(x')\rd x' = 1 .
\end{gather}
The parameter $a$ therefore specifies the length of our bar, while $b$ controls the rate of exponential decay along the major axis.

In what follows, we streamline the algebraic expressions by defining intermediate auxiliary functions to reduce clutter. Additionally, we adopt the notation that for a function $f(x, \dots)$
\begin{gather}
f_\text{s}(x) \equiv f(x) + f(-x), \qquad\qquad
f_\text{a}(x) \equiv f(x) - f(-x), \label{eq:antisymmetrisation}
\end{gather}
are the symmetrisation and antisymmetrisation respectively of $f$ with respect to $x$.

\subsection{Bar density}

The density of the bar is the convolution
\begin{gather}
\rho_{\text{bar}} = \int_{-a}^{a}w(x')\rho_{\text{Gaussian}}(x-x',y,z)\rd x' .
\end{gather}
We define a function $G(x)$ as 
\begin{gather}
G(x) = e^{-\frac{x}{b}}\left(\erf{\frac{ab-ax+\sigma^2}{\sqrt{2}b\sigma}} - \erf{\frac{-bx+\sigma^2}{\sqrt{2}b\sigma}}\right) ,
\end{gather}
with the standard error function given as 
\begin{gather}
\erf{x} = \frac{2}{\sqrt{\pi}}\int_0^x e^{-t^2} \rd t .
\end{gather}
We can compute the integral in $\rho_{\text{bar}}$ to find
\begin{gather}
\rho_{\text{bar}} = \frac{M\exp\left[\frac{2ab+\sigma^2}{2b^2}\right]}{8a(e^{a/b}-1)\pi q \sigma^2}
\exp\Big[-\frac{y^2+z^2/q^2}{2\sigma^2}\Big]G_\text{s}(x) ,
\end{gather}
where $G_\text{s}(x)$ is the symmetrisation of $G(x)$ with respect to $x$. The density profile is exactly Gaussian along the intermediate and minor axes of the triaxial figure, whilst it is exponential to a good approximation along the major axis.

\begin{figure*}
	\centering
	\includegraphics[width=.8\textwidth]{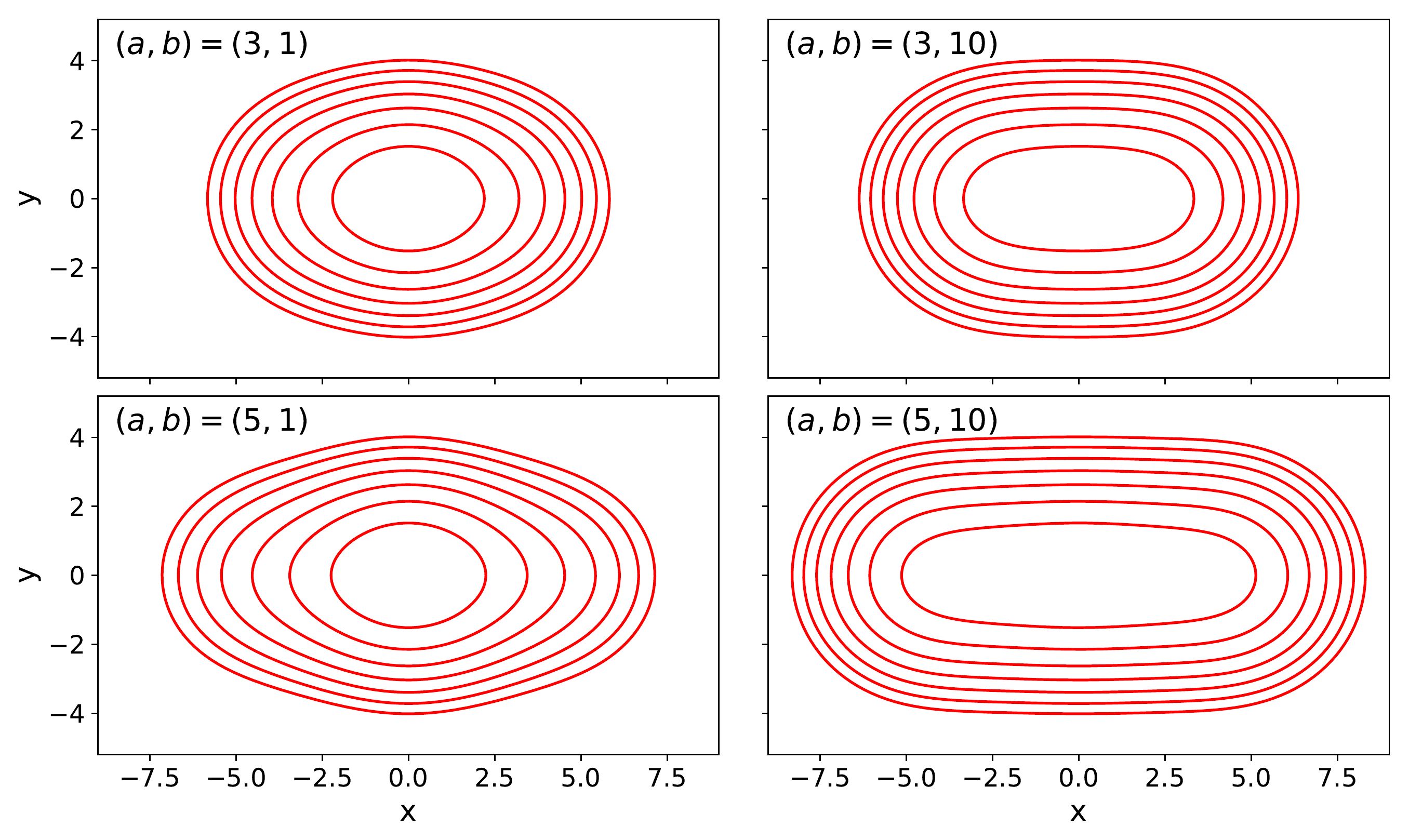} 
	\caption{Four logarithmically-spaced contour plots of $\rho$ in the $(x,y)$ plane, for varying $a$ and $b$. Top left shows $(a,b) = (3,1)$, top right $(3,10)$, bottom left $(5,1)$, bottom right $(5,10)$. It can be seen that increasing $a$ (going from the top line to the bottom) increases the length of the bar, whereas increasing $b$ (going from left to right) makes the bar flatter but does not affect the length.}
	\label{fig:rho_contour_plots}
\end{figure*}

The model bar has in total four parameters: $a$, $b$, $\sigma$ and $\epsilon$. Here, $\sigma$ is an overall length scale, which we set to unity in our plots unless otherwise stated, whilst $a$ controls the length of the bar and $b$ its flatness, as shown in the panels of Fig.~\ref{fig:rho_x_profile_2}. Together, these two parameters prescribe the shape of the density in the $(x,y)$ plane, as shown in the density contour plots of Fig.~\ref{fig:rho_contour_plots}. $\epsilon$ then controls the $z$-flattening of the bar, and has no effect on the $(x,y)$ density profiles other than scaling. We set $\epsilon = 0$ ($q=1$) unless otherwise stated.

\subsection{Surface density}

To find the surface densities, we integrate the density along a given line of sight. For simplicity, we consider the lines of sight to be the axis directions, and obtain three surface brightness functions $\Sigma_x(y,z)$, $\Sigma_y(x,z)$ and $\Sigma_z(x,y)$, namely
\begin{gather}
\Sigma_x(y,z) \propto \frac{M}{4 \pi  q \sigma ^2 \left(e^{a/b} -1\right)}\exp\left[-\frac{y^2+z^2/q^2}{2\sigma^2}\right],
\end{gather}
\begin{gather}
\Sigma_y(x,z) = \frac{M}{4b(e^{a/b}-1)\sqrt{2\pi} q \sigma^2}\exp\left[-\frac{z^2}{2q^2\sigma^2}\right]G_\text{s}(x),
\end{gather}
and
\begin{gather}
\Sigma_z(x,y) = \frac{M}{4a(e^{a/b}-1)\sqrt{2\pi} \sigma^2}\exp\left[-\frac{y^2}{2\sigma^2}\right]G_\text{s}(x).
\end{gather}
We therefore have Gaussian luminosity profiles along the minor and intermediate axes. The luminosity profile along the major axis is exponentially falling to a very good approximation (as opposed to almost constant along the bar).

\subsection{Bar potential}

To find the potential $\phi_{\text{bar}}$ for this density, we follow eqs~\eqref{eq:phi_g} and \eqref{eq:h_m}, and write
\begin{multline}
\Phi_{\text{bar}} = -\frac{GM}{\sqrt{2\pi}\sigma }\int\displaylimits_{0}^{1}\int\displaylimits_{-a}^{a}\frac{e^{-\frac{|x'|}{b}}H(x-x',y,z,m)}{b(1-e^{-a/b})} \rd x'\rd m .
\end{multline}
The inner integral can be evaluated analytically. Defining functions $F(m,x)$ and $E(m,y,z)$ by
\begin{multline}
F(m,x)=e^{-\frac{x}{b}} \left(\erf{\dfrac{bm^2(a-x)+\sigma^2}{\sqrt{2}bm\sigma}}-\erf{\dfrac{-bm^2x+\sigma^2}{\sqrt{2}bm\sigma}}\right)
\end{multline}
and
\begin{gather}
E(m,y,z)=\exp\left[-\dfrac{m^2y^2}{2\sigma^2} - \dfrac{m^2z^2}{2\sigma^2(1-m^2\epsilon^2)}+\dfrac{\sigma^2}{2b^2m^2}\right],
\end{gather}
we can write our potential as
\begin{gather}
\Phi_{\text{bar}} = -\frac{GM}{2b(1-e^{-a/b})}\int_{0}^{1}H_{\text{conv}}(x,y,z)\rd m,
\end{gather}
where
\begin{gather}
H_{\text{conv}} = \sqrt{\frac{1}{m^2(1-m^2\epsilon^2)}}E(m,y,z)F_\text{s}(m,x).
\end{gather}
The numerical computation of $H_{\text{conv}}$ needs care close to $m=0$. In order to integrate accurately, we build a Taylor approximation $H_{\text{conv}}^{\text{TS}}$ of $H_{\text{conv}}$ around $m=0$. We find a functional form for the value of $m$ at which the switchover occurs, $m=\text{Osc}(\sigma,b)$. In other words, we write our integral as 
\begin{gather}
\int\limits_{0}^{\text{Osc}(\sigma,b)}H_{\text{conv}}^{\text{TS}}(x,y,z)\rd m + \int\limits_{\text{Osc}(\sigma,b)}^1 H_{\text{conv}}(x,y,z)\rd m.
\end{gather}
To find a useful expansion, we factorise out $\exp\left[(-m^2/2\sigma^2)(R^2 + z^2/(1-\epsilon^2m^2))\right]$ (the coordinate dependence of $H$ before convolution with our weight function, see eq~\eqref{eq:h_m}) from $H_\text{conv}$ before expanding as a Taylor series. This term is $O(1)$ as $m\to 0$, and so is well-behaved. This ensures that our approximation remains valid as $|\mathbf{x}|\to\infty$.  Our Taylor approximation is
\begin{multline}
H_{\text{conv}}^{\text{TS}} = \sqrt{\frac{2}{\pi \sigma^2}}\exp\left[-\dfrac{m^2(x^2\!+\!y^2)}{2\sigma^2} - \dfrac{m^2z^2}{2\sigma^2(1\!-\!m^2\epsilon^2)}\right] \\ \bigg( 2b\left(1\!-\!e^{-a/b}\right) +  m^2\frac{b}{\sigma^2}e^{-a/b}\left(a^2 +2ab\!+\!(1\!-\!e^{a/b})(2b^2\!-\!\sigma^2\epsilon^2)\right) \bigg).
\end{multline}
The integration is performed using Gauss-Legendre quadrature \citep{Ab72} throughout. To find the onset of numerical instability as $m \rightarrow 0$, we plot $H_\text{conv}$ and vary the bar parameters continuously. We find a useful guide is to choose
\begin{gather}
\text{Osc}(\sigma,b) = 0.2\frac{\sigma}{b} . \label{eq:osc}
\end{gather}
This is a slight overestimate of the switchover point, which ensures that the results of our integration are secure.

\begin{figure}
	\centering
	\includegraphics[height=6cm]{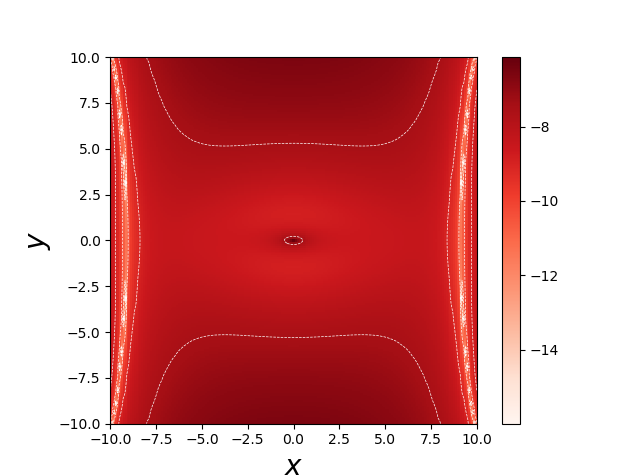} 
	\caption{Logarithm of the deviation of our force from Poisson's equation, $\log\left[\left(\nabla\cdot \mathbf{F}_\text{bar} -  -4\pi G \rho_{\text{bar}}\right)/|\vc{F}_\text{bar}|\right]$, for bar parameters $(a,b,\sigma,\epsilon) = (5,2,1,0)$. We use 20 abscissae in our numerical integration and a step size of $10^{-4}$ to evaluate derivatives with the central difference method. We note that it is everywhere very small, and so our forces are accurately computed.}
	\label{fig:deviation_from_grad_f}
\end{figure}
\begin{figure}
	\centering
    \includegraphics[width=\columnwidth]{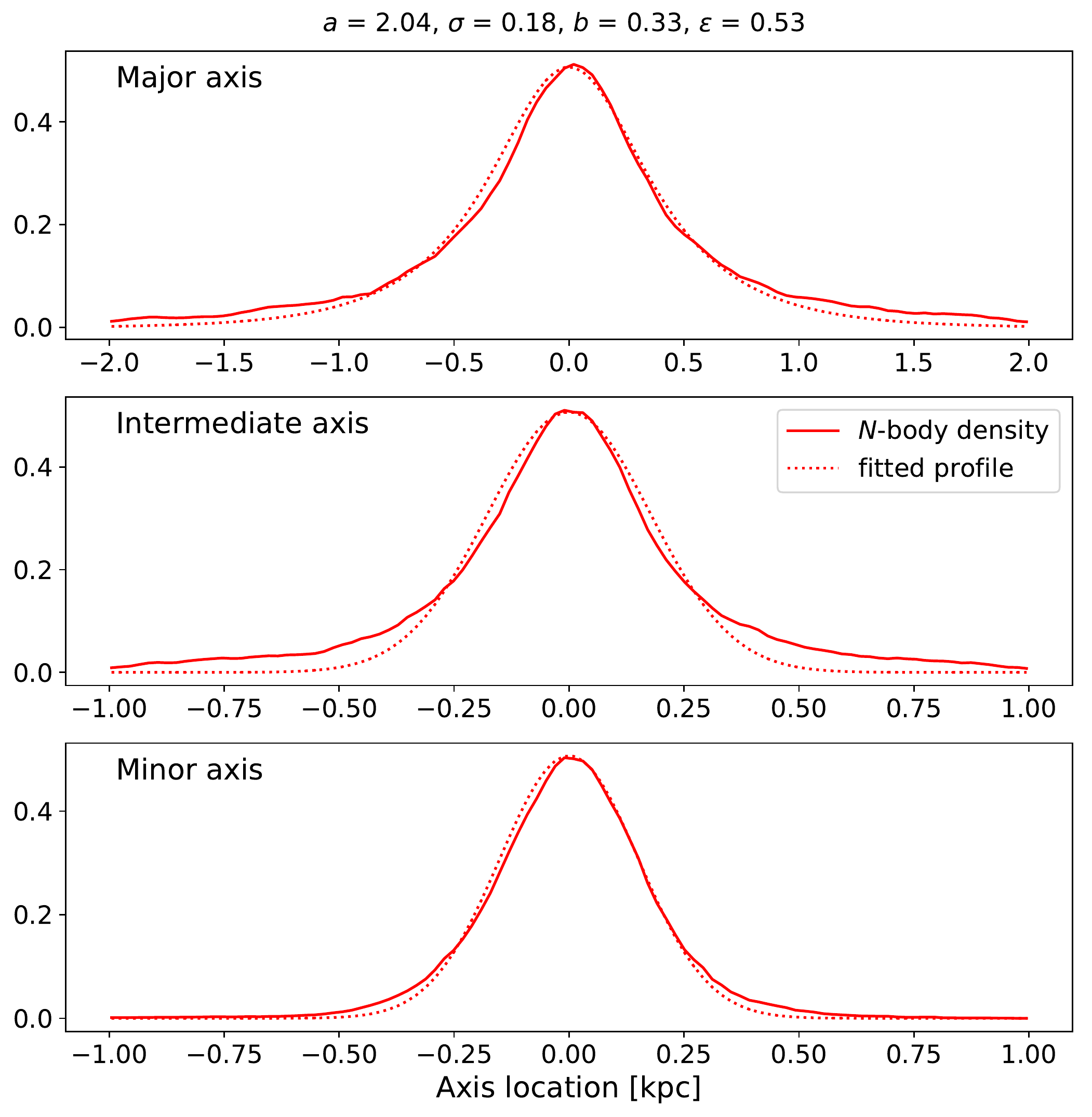}
	\caption{Major, intermediate and minor axis profiles for an $N$-body simulation and our best-fit model (which has
	the bar parameters: $(\sigma,\epsilon,a,b)= (0.1865, 0.3657\,\mathrm{kpc}, 2.8887\,\mathrm{kpc}, 0.3513\,\mathrm{kpc})$). Note the different scale for the topmost plot.
	}
	\label{fig:fitted_major_axis_profiles}
\end{figure}

\subsection{Forces}

To find the force acting on a test particle due to the bar, we need to evaluate
\begin{gather}
\mathbf{F}_{\text{bar}}= -\nabla\Phi_{\text{bar}} = \nabla\int_{0}^{1}\frac{GM}{2b(1-e^{-a/b})} H_{\text{conv}} \rd m .
\end{gather}
We take the gradient inside the integral to write
\begin{gather}
\begin{split}
\mathbf{F}_{\text{bar}} &= \frac{GM}{2b(1-e^{-a/b})}\int_{0}^{1}\nabla H_{\text{conv}} \rd m \\  &=\frac{GM}{2b(1-e^{-a/b})}\int_{0}^{1}\mathbf{F}_{\text{conv}} \rd m.
\end{split}
\end{gather}
Defining a function $I(m,x)$ by
\begin{gather}
I(m,x) = \sqrt{\frac{2}{\pi\sigma^2}}\exp\left[\dfrac{x}{b} - \dfrac{\left(bm^2(a+x)+\sigma^2\right)^2}{2b^2m^2\sigma^2}\right],
\end{gather}
we can calculate our integrand to be
\begin{gather}
\mathbf{F}_{\text{conv}} = \left(\begin{array}{c} \dfrac{E(m,y,z)}{\sqrt{1-m^2\epsilon^2}}\left[I_\text{a}(m,x)- \dfrac{1}{bm}G_\text{a}(m,x)\right]\\
-\dfrac{m^2y}{\sigma^2}H_{\text{conv}}\\
-\dfrac{m^2z}{\sigma^2(1-m^2\epsilon^2)}H_{\text{conv}}
\end{array}\right),
\end{gather}
using the notation given in eq.~\eqref{eq:antisymmetrisation}.

We encounter the same problem in evaluating components of $\mathbf{F}_{\text{conv}}$ as we faced in integrating $H_{\text{conv}}$. Again, we use a Taylor approximation, splitting our integral into two regions.  While the forces and potential are not easily computed by hand, and their analytical forms are not elegant, we can evaluate them speedily using a Gaussian quadrature, allowing for rapid investigation of the bar and its properties. 

To check the accuracy of our quadrature, we ensure that the forces satisfy Poisson's equation
\begin{gather}
\nabla\cdot \mathbf{F}_\text{bar} = - 4\pi G \rho_{\text{bar}}.
\end{gather}
Using a simple finite difference method verifies that this is the case, as shown in Fig.~\ref{fig:deviation_from_grad_f}, and so we can be confident that our integration method is faithful. Reducing $b/\sigma$ increases the proportion of the force computed using the Taylor approximation, as is evident from  eq.~\eqref{eq:osc}. This can also be a source of inaccuracy in our integration. Our numerical investigation suggests that this becomes a problem for models with $b/\sigma \lesssim 1/7$. However, this regime is unrealistic, as the ``bar" density is very spread out in the $y$ and $z$ directions but has a steeply falling $x$ profile. This is not bar-like at all, and so failure in this unphysical regime is not a serious cause for concern.

\subsection{Validation}

To ensure that our model can represent realistic bars, we match to an $N$-body simulation which is believed to mimic approximately the properties of the Milky Way. It uses the initial condition generation \texttt{mkgalaxy} from \cite{Mc07}. The simulation is described in \citet{Sa19b} and contains three components: a disc, a bulge and a dark halo (which we do not consider, as our model is for only the barred stellar component). The disc contains 200,000 particles and the bulge 40,000. The Toomre $Q$ for the disc is chosen to make a bar form rapidly.

We fit our bar model to the endpoint of this simulation. We find a set of parameters $(\sigma,\epsilon,a,b)$ that matches the profile along the three principal axes simultaneously by least squares fitting. The simulation density and our best fit are shown in Fig.~\ref{fig:fitted_major_axis_profiles}. It is clear that our model is capable of reproducing bars in $N$-body simulations well. We note that the model does underestimate the density of luminous matter at large radii. This is not surprising, as the simulation contains both bar and disc, whereas our model is
only expected to be a reasonable match to the bar.

%
\begin{figure}
	\centering
\includegraphics[width=\columnwidth]{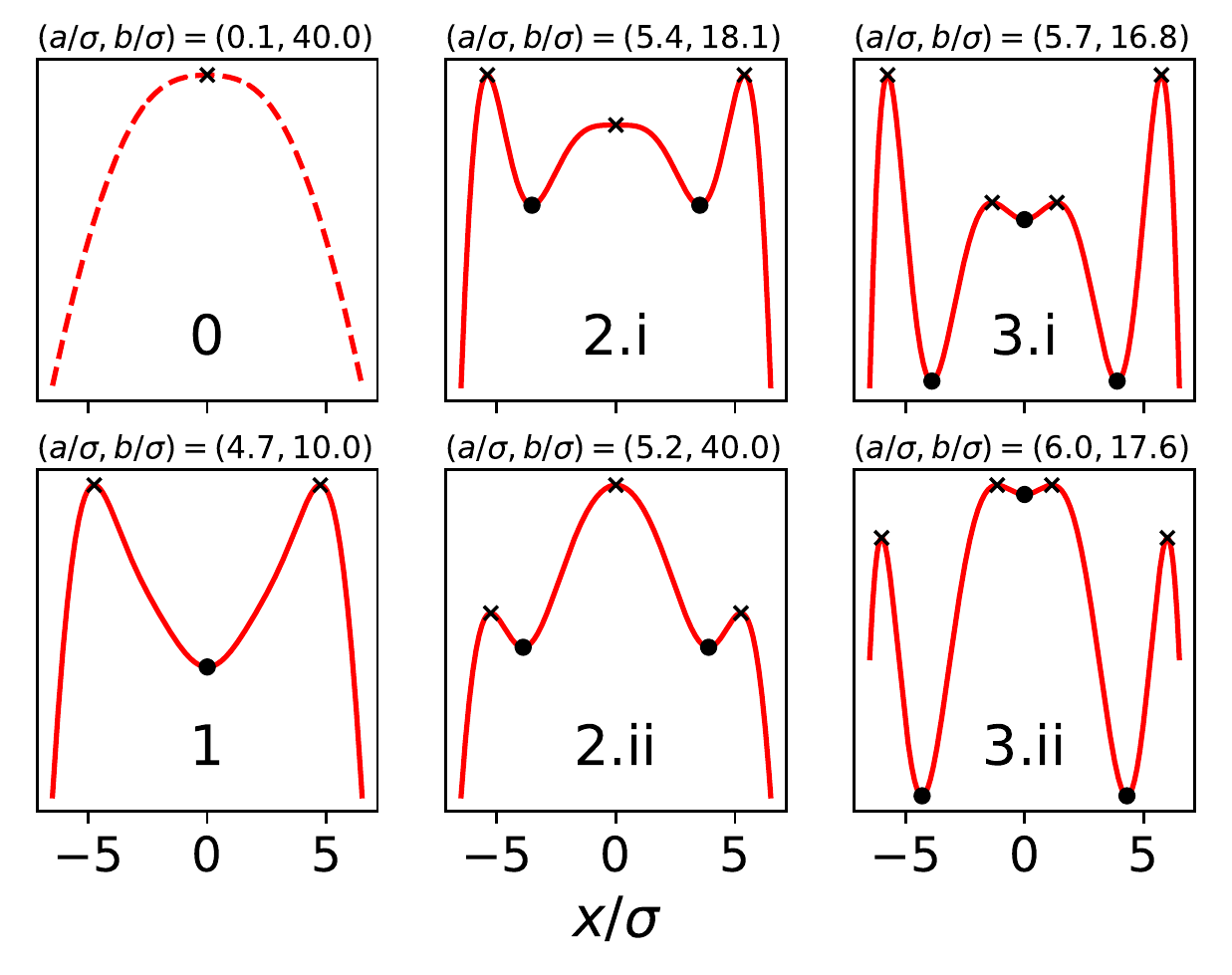} 
	\caption{\change{Schematic} plots of the effective potential $\phi_{\text{eff}}(x,0)$ against $x$ along the major axis of the bar, showing qualitatively the six types of behaviour are possible. Note that type 0 corresponds to a pattern speed too fast for any bound orbits to exist, and so is unphysical. The vertical scales vary in each image. \change{The stable and unstable fixed points are marked with circles and crosses respectively.}}
	\label{fig:eff_potential_different_types}
\end{figure}
\begin{figure}
	\centering
	\includegraphics[height=6cm]{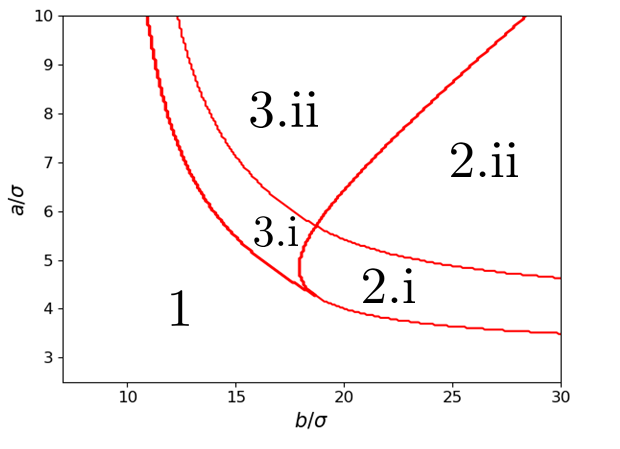} 
	\caption{Regions of $(b/\sigma,a/\sigma)$ parameter space in which each type of effective potential can be found.}
	\label{fig:eff_potential_parameter_space_a}
\end{figure}
\begin{figure*}
	\centering
	\includegraphics[height=8cm]{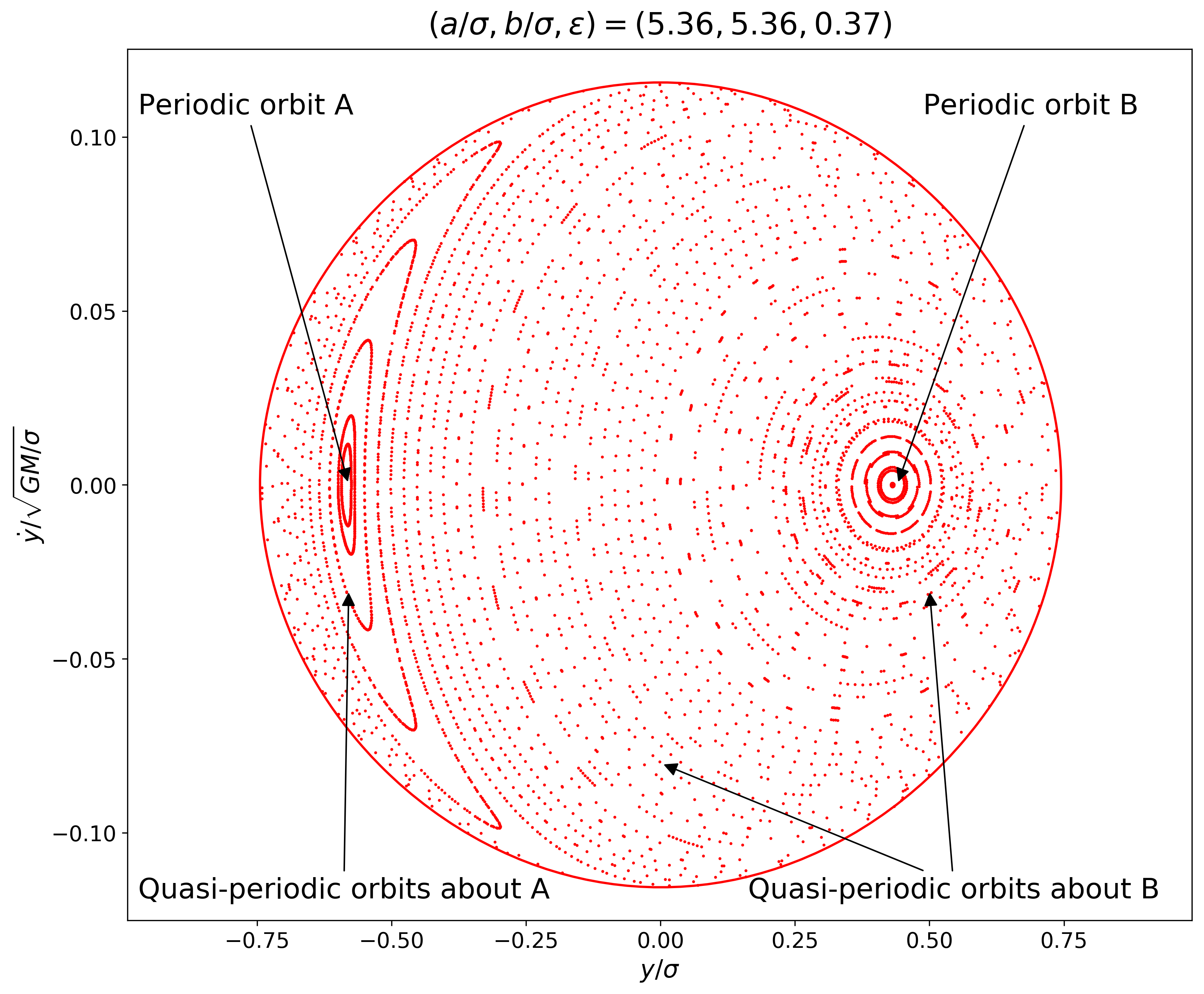} 
	\caption{Poincar\'e surface of section for a type 1 potential (single potential well), with $a=1$ and $b=1$ at energy $H_J = -2.5$. \change{Two distinct periodic orbits can be seen -- they are labelled as A (which sires the $x_4$ family) and B (which sires the $x_1$ family). These labels correspond to the orbits shown in Figure \ref{fig:periodic_orbits_for_section_1}. The quasi-periodic orbits librating around the stable orbits map out the invariant curves around each fixed point. All displayed orbits have energy conservation better than $|\Delta E/E|=10^{-8.5}$.}}
	\label{fig:poincare_section_1}
\end{figure*}
\begin{figure*}
	\includegraphics[height=6cm]{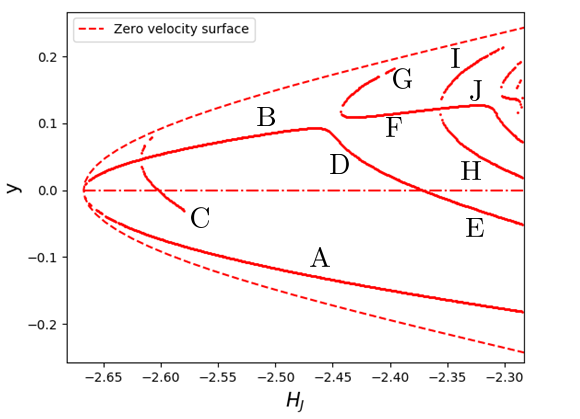} 
    \includegraphics[height=6cm]{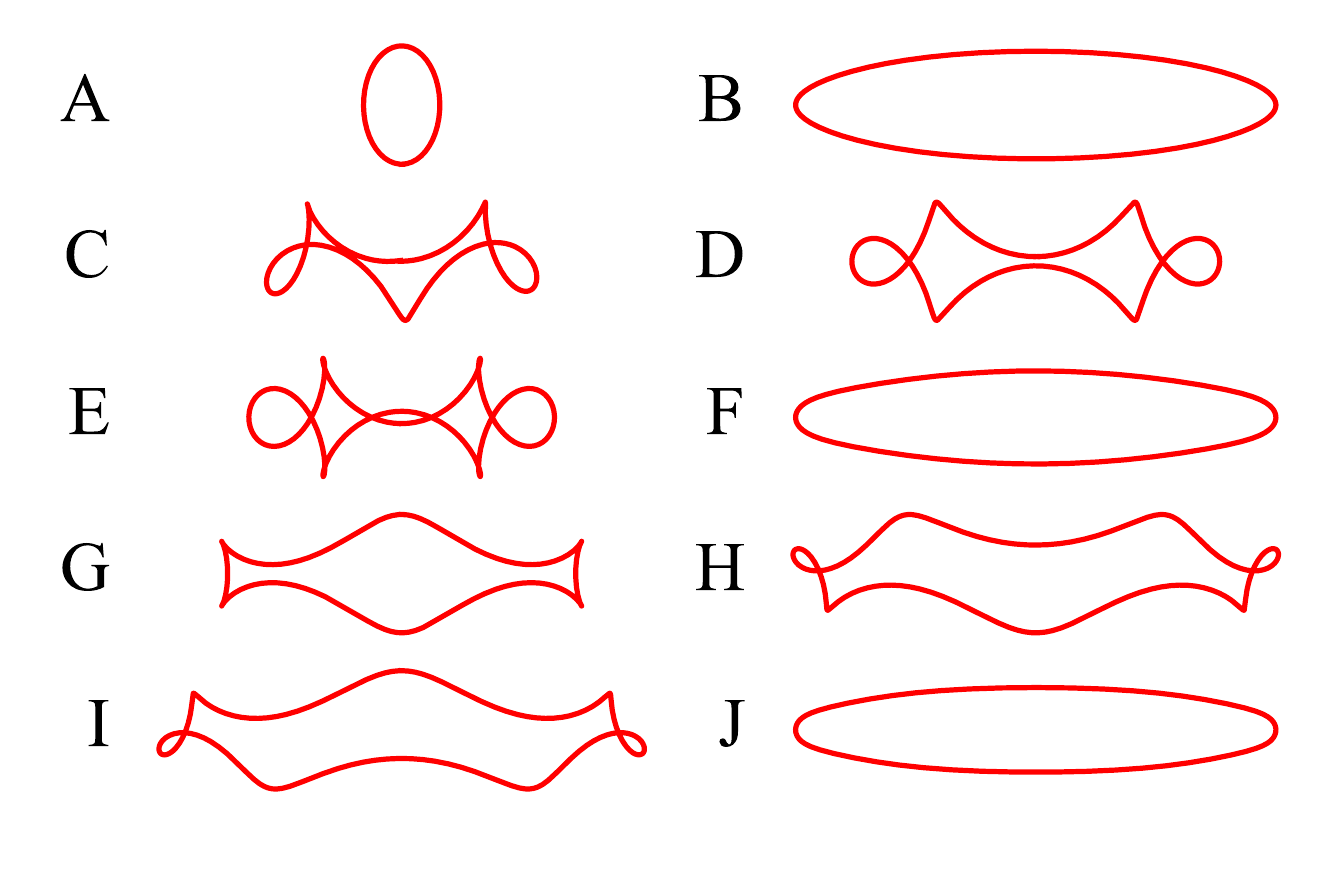}
	\caption{Left: \change{Characteristic diagram showing where periodic orbits with energy $H_J$ intersect the $x=0$ plane. Each periodic orbit is labelled with a capital letter. Orbits with $y>0$ are prograde (rotating with the rotation of the bar), whereas orbits with $y<0$ are retrograde. Right: The periodic orbits present in our bar for $a=1,b=1,\sigma=0.2,\epsilon=0.4$.
	Orbits B, F, J are the $x_1$ backbone of the bar, with other orbits bifurcating away from this.}}
	\label{fig:periodic_orbits_for_section_1}
\end{figure*}
\section{Orbital structure}\label{section:eff_pot_structure}

With our expressions for the forces of our new model, we now turn to inspecting the orbital structure for different parameter choices.

\subsection{The Effective Potential}

We restrict our attention to orbits with $z=0$. As the bar rotates, we transform to a frame of reference which corotates with the bar clockwise at a constant pattern speed of $\Omega_b$. This is consistent with standard convention for the Milky Way's bar. The equations of motion in this frame are \citep[e.g.,][]{BT}:
\begin{gather}
\ddot{\mathbf{r}} = -\nabla \phi - 2(\mathbf{\Omega}_b \times \dot{\mathbf{r}}) - \mathbf{\Omega}_b \times (\mathbf{\Omega}_b \times \mathbf{r}),
\end{gather}
where $\mathbf{\Omega}_b = -\Omega_b\mathbf{e}_z$, with a negative sign ensuring clockwise rotation when in a right-handed coordinate frame. 
%
We can view this as motion in an effective potential:
\begin{gather}
\ddot{\mathbf{r}} = -\nabla \Phi_\text{eff} - 2(\mathbf{\Omega}_b \times \dot{\mathbf{r}}) ,
\end{gather}
where $\Phi_\text{eff}$ is
\begin{gather}
\Phi_{\text{eff}} = \Phi_{\text{bar}} - \frac{1}{2}\Omega_b^2(x^2 + y^2).
\end{gather}
The quantity $H_J = \frac{1}{2}|\dot{\mathbf{r}}|^2 + \phi_{\text{eff}}$, known as the Jacobi integral or the energy in the rotating frame, is conserved.

Corotation occurs near the end of the bar, which in our case is $x=a$. We can use this to fix the pattern speed as \citep[c.f.][]{Wi17}
\begin{gather}
\Omega_b = \sqrt{\frac{-F_x(a,0)}{a}}.
\end{gather}
Each combination of parameters $\left(\sigma,\epsilon,a,b\right)$ gives a distinct bar, with considerable scope for different behaviour as these parameters vary. In what follows, we fix $\sigma = 0.1865$ and $\epsilon = 0.3657$ (our best fit values in Fig.~\ref{fig:fitted_major_axis_profiles}) unless otherwise specified, and treat $a$ and $b$ as independent parameters. \change{A useful discriminant is the profile of the effective potential shape, and hence the number of stable and unstable Lagrange points, on the major axis. There are six different types of possible profile, as shown in Fig.~\ref{fig:eff_potential_different_types} and categorized below:}
\begin{itemize}
	\item[0] - Here the pattern speed is too large for any \change{bound} orbits to exist,
	\item [1]- The origin is the only stable fixed point,
	\item [2] - The origin is unstable, but two stable fixed points appear away from the origin,
	\begin{itemize}
		\item [2.i] - The effective potential is small enough at the origin that orbits can encircle both stable fixed points,
		\item [2.ii] - No orbits can encircle both fixed points, and orbits originating at the origin are unbound,
	\end{itemize}
	\item [3] - The origin and two other fixed points are all stable,
	\begin{itemize}
		\item [3.i] - Orbits can exist encircling all three fixed points, as the maximum of the effective potential is outside the three fixed points,
		\item [3.ii] - The three stable fixed points are isolated - orbits with enough energy to travel from the origin to the other fixed points are unbound.
	\end{itemize}
\end{itemize}
In Fig.~\ref{fig:eff_potential_parameter_space_a}, we label the regions in $(b/\sigma,a/\sigma)$ parameter space in which each type of effective potential can be found. We note also the effect of varying $\epsilon$ on our configuration space. Increasing $\epsilon$ corresponds to making the bar's mass more concentrated in the $(x,y)$ plane. This causes Region 1 grow at the expense of the other regions.

Types 1 and 2.i are seen in many types of bar, such as the numerical Cazes bars \citep{Ba01}, and the prolate $n = 2$ Ferrers ellipsoid supplemented by a Plummer sphere \citep{Ka05}. \change{They show the standard picture of 3 unstable Lagrange points (L$_1$, L$_2$ and L$_3$) on the major axis, interleaved with 2 stable ones~\citep[e.g.,][]{Wi17}.}

A key observation is that type 3 potentials (the right hand panels of Figure \ref{fig:eff_potential_different_types}) are not seen in the \citet{Wi17} flat \lq\lq logarithmic bar", and appear to be an unusual property of this model. \change{There are now 4 unstable Lagrange points and 3 stable ones on the major axis.) In fact, if we fix $a$ and increase $b$ to make a flattish bar of a fixed length, type 3 potentials no longer exist, and we regain the types of potential seen for the flat bar. This suggests that type 3 potentials are a feature present only in bars with a rapidly falling density profile along the major axis. There are a few other models in the literature that exhibit additional Lagrange points on the major axis, including the asymmetric bar shown in Fig 2 of \citet{Pa10} and the inhomogeneous bar of Fig 9 of \citet{At09}.}

\begin{figure}
	\centering
	\includegraphics[height=6cm]{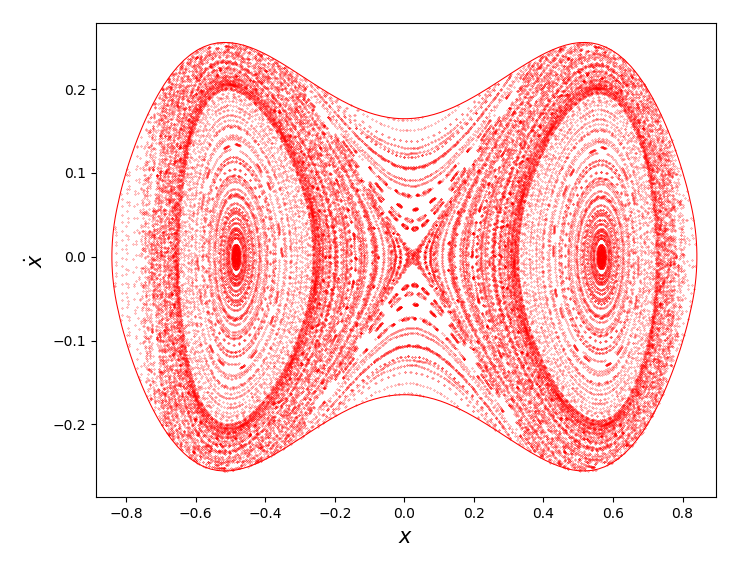} 
	\caption{\change{Poincar\'e section for a typical type 2 potential, with $a=0.9$, $b=5$, $\sigma=0.2$ and $\epsilon=0.4$ at energy $H_J = -2.46$. 
	$x_4$-like periodic orbits can be seen at the two stable Lagrange points (or $x_4'$ and $x_4''$ orbits in the language of \protect\cite{Pa10}). Some orbits circulate around both Lagrange points, shown by the invariant curves encircling the entire section. There is a thin chaotic layer associated with the separatrix, which emanates from the unstable Lagrange point. The values of the effective potential at the Lagrange points are $-2.457,-2.493^*,-2.474$ where ${}^*$ denotes if stable. This plot may be usefully compared with Figure 7 of \citet{Wi17} for flattish bars}.}
	\label{fig:poincare_section_2i}
\end{figure}
\begin{figure}
	\centering
	\includegraphics[height=6cm]{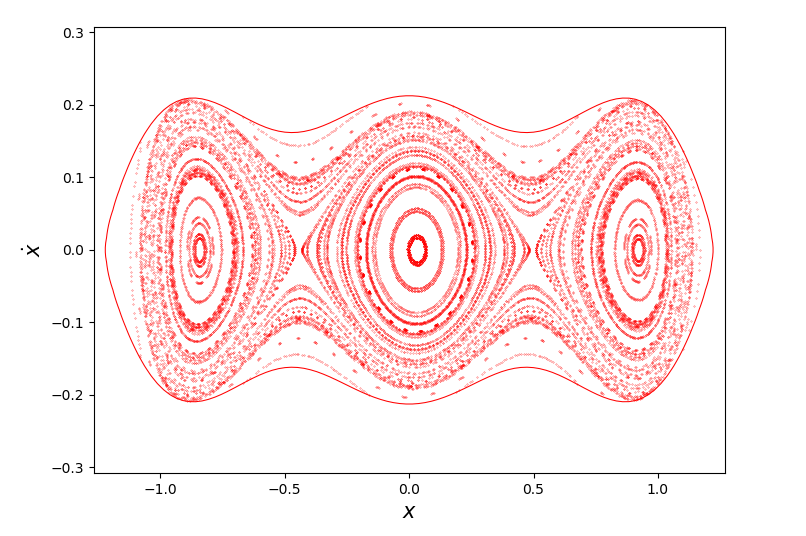} 
	\caption{\change{Poincar\'e section for a typical type 3 potential, with $a=1.25$, $b=3$, $\sigma=0.2$ and $\epsilon=0.4$ at energy $H_J = -2.111$. 
	There are now 3 stable Lagrange points with quasi-periodic orbits encircling them. Some orbits circulate around all three stable Lagrange points. The values of the effective potential at the Lagrange points are $-2.110,-2.133^*,-2.124,-2.134^*$ where ${}^*$ denotes if stable}.}
	\label{fig:poincare_section_3i}
\end{figure}
\begin{figure*}
	\centering
	\includegraphics[height=10cm]{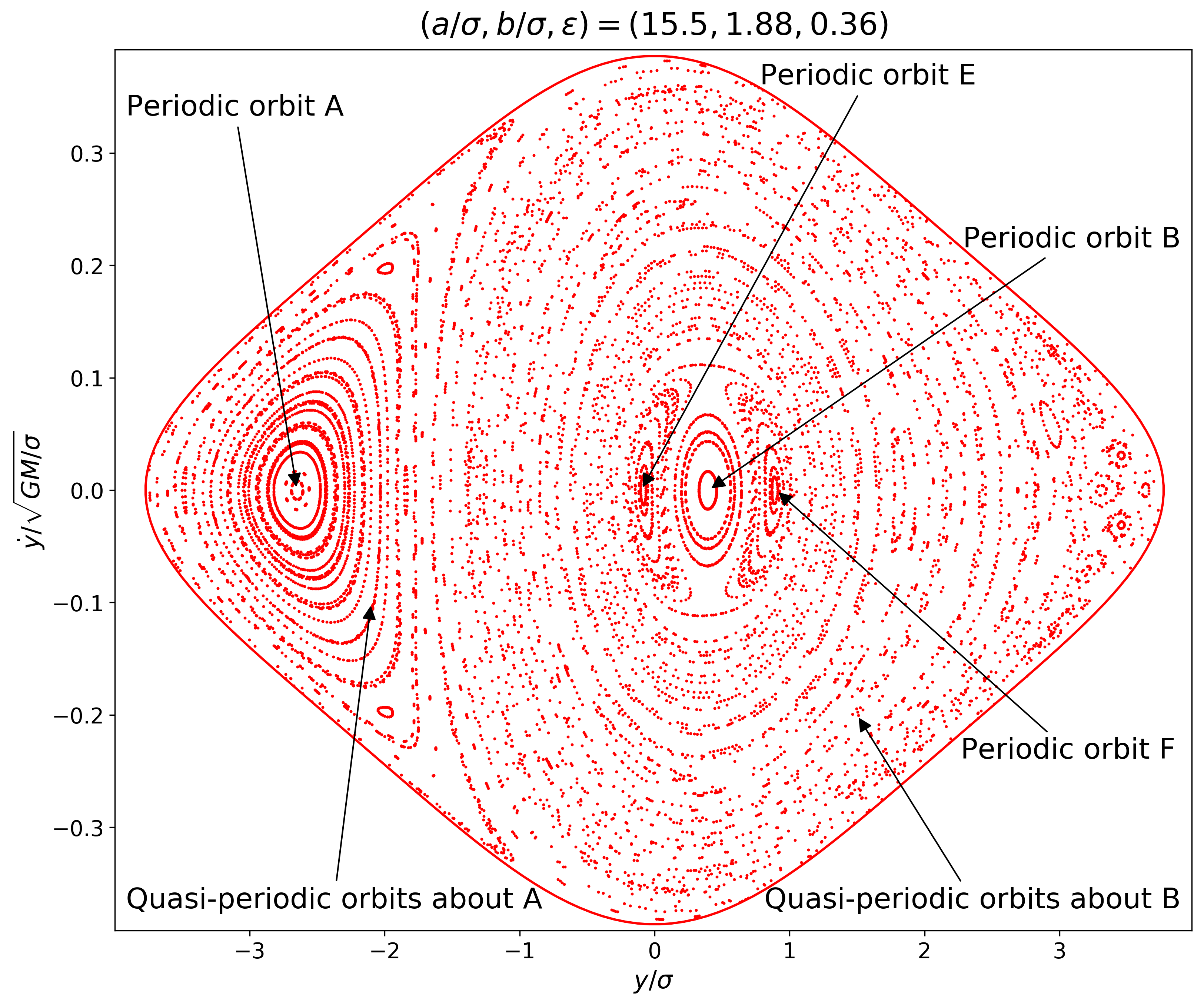}
	\caption{Poincar\'e surface of section for our exponential bar, with $a=2.89$ and $b=0.35$ at energy $H_J = -1.2$. 4 periodic orbits can be seen, at $y\approx -0.24,-0.15,-0.03$ and $0.6$, and are labelled, along with \change{the quasi-periodic orbits that they sire}.}
	\label{fig:poincare_section_exp_bar}
\end{figure*}
\begin{figure*}
    \begin{minipage}[l]{0.4\textwidth}
	\includegraphics[height=5.5cm]{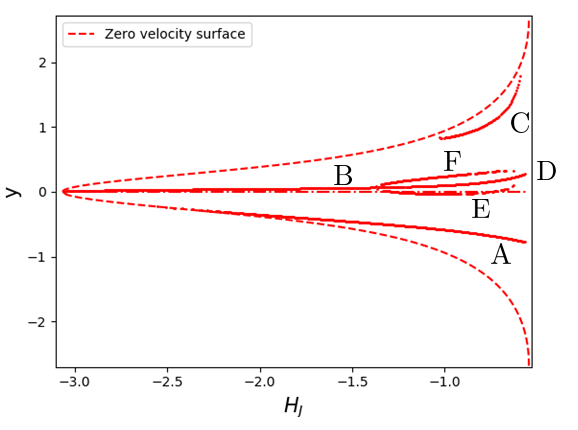}
	\end{minipage}
	\begin{minipage}[r]{0.4\textwidth}
	\includegraphics[height=5.5cm]{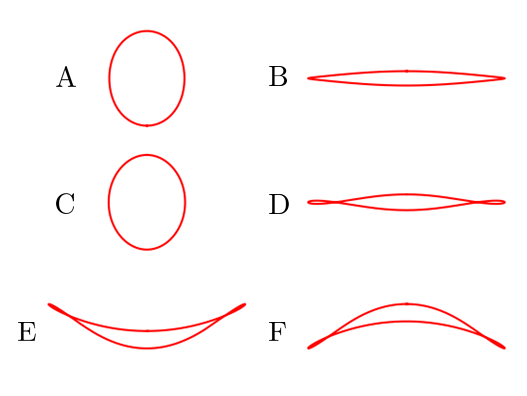}
	\end{minipage}
	\caption{Left: Periodic orbit structure for the exponential bar. Orbits are again labelled with capital letter. Orbits with $y>0$ are prograde, whereas orbits with $y<0$ are retrograde. The zero velocity surface is given by $H_J = \phi_{\text{eff}}$, such that orbits on this surface have zero kinetic energy.
	Right: The periodic orbits present in our bar for $a=2.89,b=0.35$.
	}
	\label{fig:periodic_orbits_for_exp_bar}
\end{figure*}

\subsection{Poincar\'e sections and orbital structure}\label{section:poincare}

\change{A valuable tool, which has been used extensively to visualise the orbital structure in bar potentials, is the Poincar\'e surface of section \citep[e.g.,][]{He64,RCD,Co02,BT}}. Our phase space for planar orbits is four dimensional: $(x,y,\dot{x},\dot{y})$. The conservation of energy reduces our phase space to three dimensions, which remains difficult to visualise and interpret. We therefore choose points $(y,\dot{y})$ along our orbit with $x=0$ and $\dot{x} > 0$, and plot them on a plane. This image is the ``Poincar\'e section". It gives a snapshot of the phase space at fixed energy. We can visualise this as setting up a photographic plate in our phase space, and letting a trajectory make a mark on the plate as it passes through in the right direction. Poincar\'e sections reveal the structure of the phase space at a fixed energy, and allow us to find periodic orbits. 

The Poincar\'e return map, $P: S\to S$, maps a point on the surface of section $S$ to the next point at which its orbit will intersect the surface: $P(\mathbf{x}_0) = \mathbf{x}_1$. \change{It is Hamiltonian and area preserving~\citep[e.g.,][]{RCD,Co02}}. This means its Jacobian has determinant equal to unity. \change{If a periodic orbit is stable, then its two eigenvalues are complex and lie on the unit circle~\citep{He65}}. This means a nearby orbit librates around the periodic orbit.  For an unstable periodic orbit, the eigenvalues are real and one lies outside the unit circle. Those orbits starting nearby will therefore quickly diverge away. We are particularly interested in the stable periodic orbits -- they will sire orbital families, which librate around them.

\change{To integrate the orbits, we use two methods: a fourth order Runge-Kutta integration method with a fixed timestep, or, where more accuracy is required, the adaptive time-stepping routine LSODA in scipy.} We integrate over 200 time units, corresponding to approximately 1 Gyr for the Milky Way-like bar in Section \ref{section:hvpeaks}. We test the accuracy of our orbits by ensuring that the energy $H_J$ is conserved for a random orbit sample -- this is true to one part in $10^{-6}$ in most cases \change{(the energy conservation deteriorates for orbits that are marginally bound and move out to very large radius)}. For situations that require an accurate intersection of a trajectory with a surface, such as the Poincar\'e sections later in this Section, we use a secondary integration to refine our solution in a given interval. \change{An example is shown in Fig.~\ref{fig:poincare_section_1} where all displayed orbits have energy conservation better than $|\Delta E/E|=10^{-8.5}$.} To calculate the force, we \change{typically} use 20 abscissae in our Gauss-Legendre quadrature. Increasing the number of abscissae from 20 to 200 reduces the error in conservation of energy by roughly a factor of \change{40}. \change{Further improvements are possible by increasing the order of the Taylor expansion employed and in turn increasing the factor used in $\mathrm{Osc}(\sigma,b)$ slightly.}

\subsubsection{Type 1 potentials}

We first investigate a short, flattish bar with $a=1$ and $b=1$ and compare this to the bars investigated by \citet{Wi17}. In Fig.~\ref{fig:poincare_section_1}, we see two different periodic orbits at $y \approx -0.11$ and $0.09$. \change{The invariant curves enclosing the fixed points on the Poincar\'e section are the quasi-periodic orbits that librate around the periodic orbits.} A single surface of section is only a snapshot of the phase space at a fixed energy. To build a full picture, we search for periodic orbits as fixed points of the Poincar\'e map $P$, and plot their intersection $y$ coordinate against the energy at which they are present. This is known as a characteristic diagram~\citep[e.g.,][]{Co02,Ka05,Wi17}. Fig.~\ref{fig:periodic_orbits_for_section_1} shows the characteristic diagram for the bar with $a=1$ and $b=1$. We can see a variety of different orbital families emerging 
\change{as the periodic families bifurcate and connect onto either the zero-velocity surface or another family (some tracks are artificially truncated in the figure due to the resolution of our periodic orbit search)}.
Each family is labelled with a letter --  from A through to J in this instance. Our phase space is dominated by the $x_1$ (elongated prograde) and $x_4$ (retrograde) orbits, using the nomenclature of \cite{Co02}. The $x_1$ sequence is B $\to$ F $\to$ J, with successive orbits bifurcating away from the main sequence. The $x_4$ orbit is A. It sires orbits that are retrograde and so unlikely to be highly populated in a real bar, as significant counter-streaming is not generally observed~\citep[e.g.][]{Se93,Ka05}. \change{Although the majority of the phase space in Fig.~\ref{fig:poincare_section_1} is indeed taken up by the $x_1$ family, this does not by itself ensure that such orbits are populated. For example, in Figure 21 of \citet{Pa97}, the area occupied by $x_4$ orbits is much greater than $x_1$, even though any actual bar must be dominated by the prograde orbits.}

\change{The phenomenon of $x_1$ orbits producing orbits through successive bifurcations has been witnessed before many times in bar models~\citep[e.g.,][]{Co89,Co02}. This is a generic feature of rotating Hamiltonians, driven by the existence of resonances. Another very common feature visible in Fig.~\ref{fig:poincare_section_1} is the co-existence of the bifurcating $x_1$ sequence with a stable $x_4$ orbit (A). This too has been seen before, first by \citet{Pf84a} and subsequently by others \citep{Co02, Ka05, Co13}}.

\subsubsection{Type 2 and 3 potentials}

We now focus on type 2.i and 3.1 potentials. Of course, types 2.ii and 3.ii are unable to be self-consistent or resemble real bars, as bound orbits do not exist. We take as our type 2.i potential $(a,b) = (0.9,5)$, and as our type 3.i potential $(a,b) = (1.25,3)$.

As all orbits intersect the line $y=0$, we construct our Poincar\'e section by looking at intersection with the $(x,\dot{x})$ plane. The range of energies for bound orbits is much smaller for type 2 and 3 potentials. The orbital structure shown in Figs.~\ref{fig:poincare_section_2i} and \ref{fig:poincare_section_3i} is generic: $x_4$ orbits around each Lagrange point, each taking up a roughly equal portion of phase space, with no other major orbital families \change{(\cite{Pa10} dub these off-centre $x_4$ orbits, $x_4$-like or $x_4'$ and $x_4''$)}. This behaviour is similar to the ``double well" potential investigated in \cite{Wi17}.

\subsection{Orbital structure in a Milky Way-like bar}

We now investigate the bar that closely resembles the simulation of \citet{Sa19b}, which itself is a reasonable match to the Milky Way's bar. We take the bar parameters $(a,b) = (2.89,0.35)$, which corresponds to a long, bar with an exponential profile. We first show a representative Poincar\'e section in Fig.~\ref{fig:poincare_section_exp_bar}. We see that there are portions of phase space in which the invariant curves have broken up and chaotic orbits occur and no discernible invariant curves can be seen. 

The characteristic diagram is shown in Figure \ref{fig:periodic_orbits_for_exp_bar}.
The orbital structure here is markedly different. We no longer have a prograde $x_1$ orbit and a retrograde $x_4$ orbit present at all energies. Instead, the dominant orbital family is a propeller orbit, B \citep{Ka05,Wi17}. These are so called because they are long and thin, and can appear as an elongated figure 8 shape, similar to a propeller. This orbit only undergoes one bifurcation, giving two offshoots, and persists at all energies. This propeller family coexists with an $x_4$ orbit A, with an $x_1$ orbit only appearing at high energies (C). The dominance of the propeller orbits in this analogue of the Milky Way bar suggests that this orbital family may be responsible for its thinness and morphology.

The orbital structure is in fact strikingly similar to the ``Model 6" bar in \cite{Ka05} and \cite{Ka96}. This bar was constructed as a best fit to NGC 1073, an SB(rs)c galaxy possessing an exponential bar with half-length 2.95 kpc. Their model consists of a $n=2$ Ferrers ellipsoid, exponential disc and a spiral perturbation. This suggests that propeller orbits are a generic feature of many galactic bars, not just the Milky Way's. 

\section{The High Velocity Peaks}\label{section:hvpeaks}

\change{\cite{Ni12} commissioned observations with the Apache Point Observatory Galactic Evolution Experiment (APOGEE), and obtained radial velocities for $\sim$4700 stars. They observed distinctive peaks in these observations, offset from the main peak by $\sim$200\, kms$^{-1}$. Here, we use our bar model to identify a new orbital family that may be responsible for the peaks. In the era of Gaia, we have access to proper motions of stars in these high velocity peaks which we use to check the validity of our proposed orbital family.}
 
\subsection{Observations}

\change{In \cite{Ni12}, radial velocities for stars in the Milky Way bulge were acquired using infra-red spectroscopy, at several different Galactic longitudes and latitudes $(l,b)$. Histograms of these radial velocities revealed high velocity peaks -- an asymmetrical feature in which approximately 10\% of stars have markedly higher radial velocities than the majority of those observed. They fit these observed velocities to a double Gaussian shape, with the parameters in Table \ref{tab:nidever_observations}. Although the peaks are also observed at fields a few degrees above and below Galactic plane, we concentrate here on the two $b=0^\circ$ fields, for which the preceding orbital analysis is pertinent.}

\change{\citet{Ni12} argue that the peaks are unlikely to be due to stochastic effects, due to the observation of the same feature in several different observation fields, though this has been contested by \citet{Li14}. Furthermore, \citet{Ni12} rule out attributing the peaks to the Sagittarius dwarf galaxy, or to undiscovered substructure in the Milky Way. They conclude that ``the best explanation for the high-velocity stars is that they are members of the Milky Way bar, grouped on particular orbits". This suggestion was made plausible by \citet{Mo15} and \citet{Au15}, who separately identified candidate families in $N$-body bars capable of reproducing the peaks. A related hypothesis is that the peaks are caused by stars moving on orbits in a kiloparsec-sized nuclear disc~\citep{De15}.}

\begin{figure}
	\centering
	\includegraphics[height=7cm]{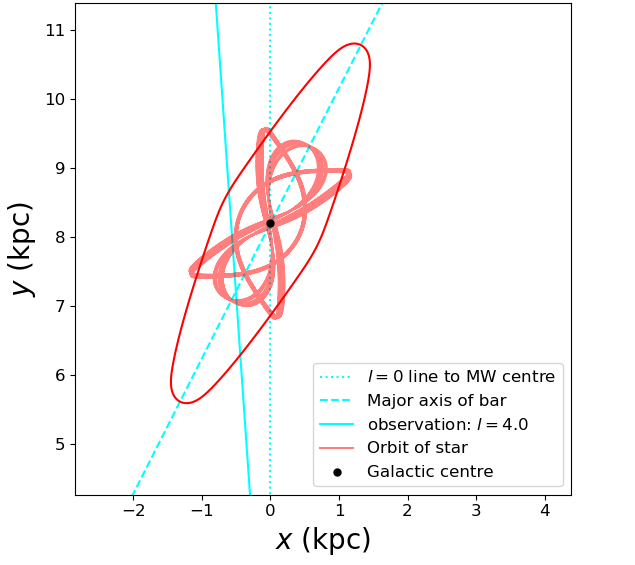} 
	\caption{Our viewpoint of the orbits in the bar potential from the Sun (situated at the origin). The outline of the bar, given by $\rho = \rho(x=a,y=0)$, is shown for reference. Here $(l,b)$ are standard Galactic coordinates. The bar here is angled at $27^\circ$ to the line $l=0^\circ$, and we observe along the line $l=4^\circ$.}
	\label{fig:solar_pov_orbit}
\end{figure}

\begin{table}
	\centering
	\begin{tabular}{l|l|l|l|l|}
		\cline{2-5}
		& \multicolumn{2}{l|}{Main stars}             & \multicolumn{2}{l|}{High velocity stars}      \\ \cline{2-5} 
		& $\left\langle V_\text{gsr} \right\rangle$ & $\sigma_v$ & $\left\langle V_\text{gsr} \right\rangle$ & $\sigma_v$   \\ \hline
		\multicolumn{1}{|l|}{$(l,b) = (4^\circ,0^\circ)$} & $ 39.4$                        & $100.9$    & $203.6$                        & $\approx 30$ \\ \hline
		\multicolumn{1}{|l|}{$(l,b) = (6^\circ,0^\circ)$} & $50.2$                         & $96.1$     & $234.3$                        & $\approx 30$ \\ \hline
	\end{tabular}
	\caption{Selection of observations of high velocity peaks, listing mean and variance of the double Gaussian fit, in units of kms$^{-1}$.}
	\label{tab:nidever_observations}
\end{table}
\subsection{Galactic geometry and coordinate transformations}

We aim to find candidate orbital families responsible for these high velocity peaks, using our exponential bar with parameters $(\sigma,\epsilon,a,b) = (0.1865, 0.3657, 2.8887, 0.3513)$, which is the best fit to the Milky Way simulation from \cite{Sa19b}. We integrate orbits in the frame of reference corotating with the bar using a pattern speed $\Omega_b$ of 40 km s$^{-1}$kpc$^{-1}$ ~\citep{Sa19b,Bo19}, and perform a coordinate transformation so that we have the orbital position and velocity from the point of view of the Sun. We then find each time the orbit crosses our line of sight, and extract the radial and tangential velocities at this point. In the heliocentric frame, the positions and velocities are
\begin{gather}
\vc{x}= \vc{R}\vc{x}' +  \left(\begin{array}{cc} 0 \\ d \\ \end{array}\right),\qquad\qquad
\vc{v} = \vc{R}\vc{v}' + \Omega_b R \hat{\vc{e}}_\theta + \vc{v}_\odot.
\end{gather}
where $R$ is the distance of the star from the centre of the bar, $\hat{\vc{e}}_\theta$ is the unit vector in the direction of increasing $\theta$, and $\vc{v}_\odot$ is the motion of the Sun in relation to the Galactic centre. The matrix $\vc{R}$
%
%
is a rotation counterclockwise by $90^\circ-\alpha$. The radial and transverse velocities are then
\begin{gather}
\begin{array}{l c l}
v_\text{rad} &=& -v_x \sin \ell + v_y \cos \ell \\
v_\text{trans} &=& -v_x \cos \ell - v_y \sin \ell .
\end{array}
\end{gather}
We can extract the proper motion $\mu_l$, measured in milliarcseconds per year, from the transverse velocity using $\mu_l = v_\text{trans}/(4.74s)$, where $s$ is the heliocentric distance. We also work with the radial velocity corrected for the solar motion: $V_\mathrm{gsr}$.

\begin{figure}
	\centering
	\includegraphics[width=\columnwidth]{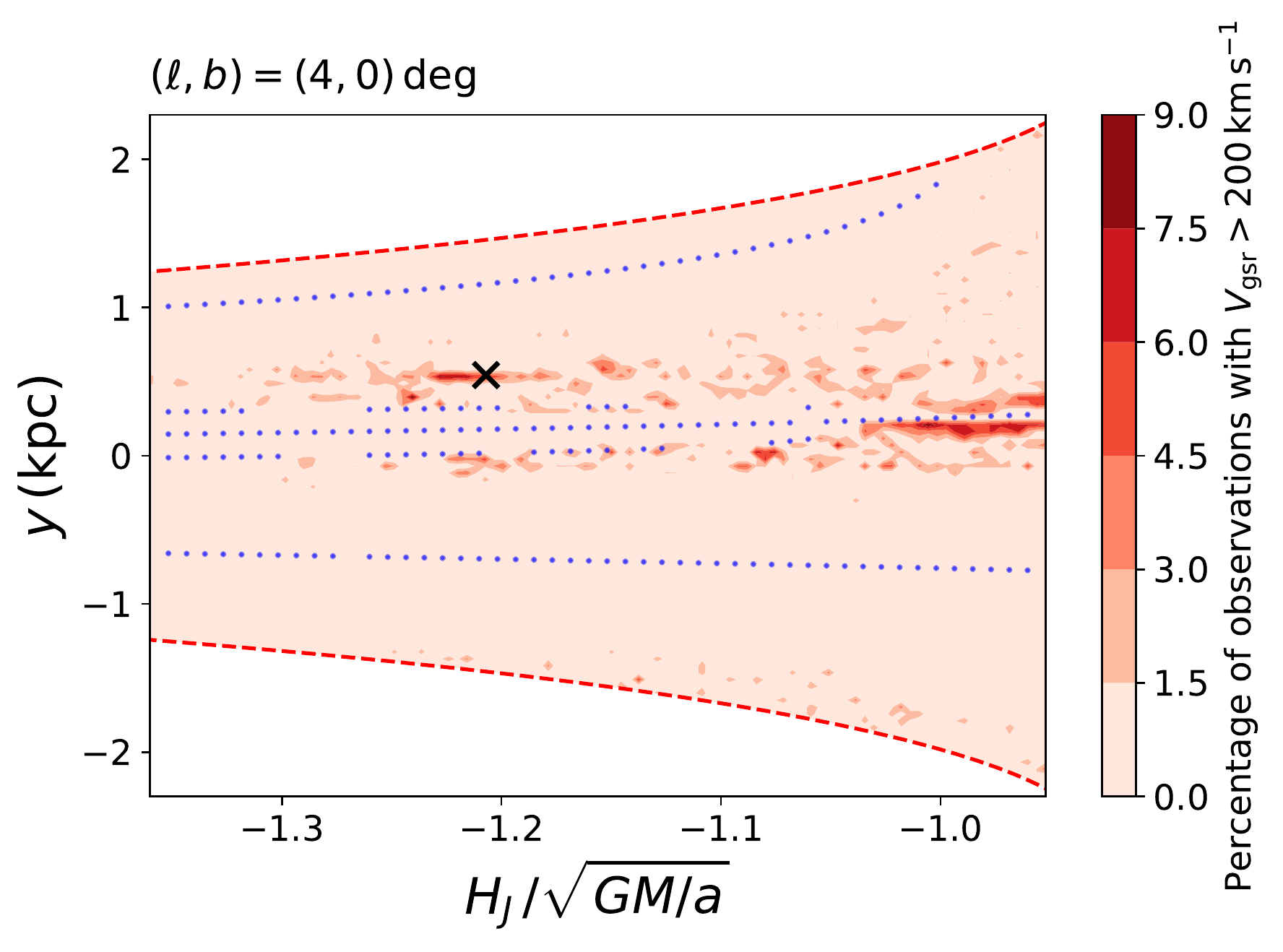}
	\caption{Percentage of observed radial velocities greater than $200\,\mathrm{km\,s}^{-1}$ for trajectories with given initial $y$ coordinate and energy, with dark regions corresponding to a high proportion of high velocity observations. Periodic orbit families are overlaid for reference. Only the central propeller family and its sirings produce such high radial velocities. The black cross marks the orbit inspected in detail in Fig.~\ref{fig:velocity_histogram}.}
	\label{fig:high_velocity_proportion}
\end{figure}
\begin{figure*}
	\centering
    \begin{minipage}{0.2\textwidth}
        \centering
    \includegraphics[height=6.0cm]{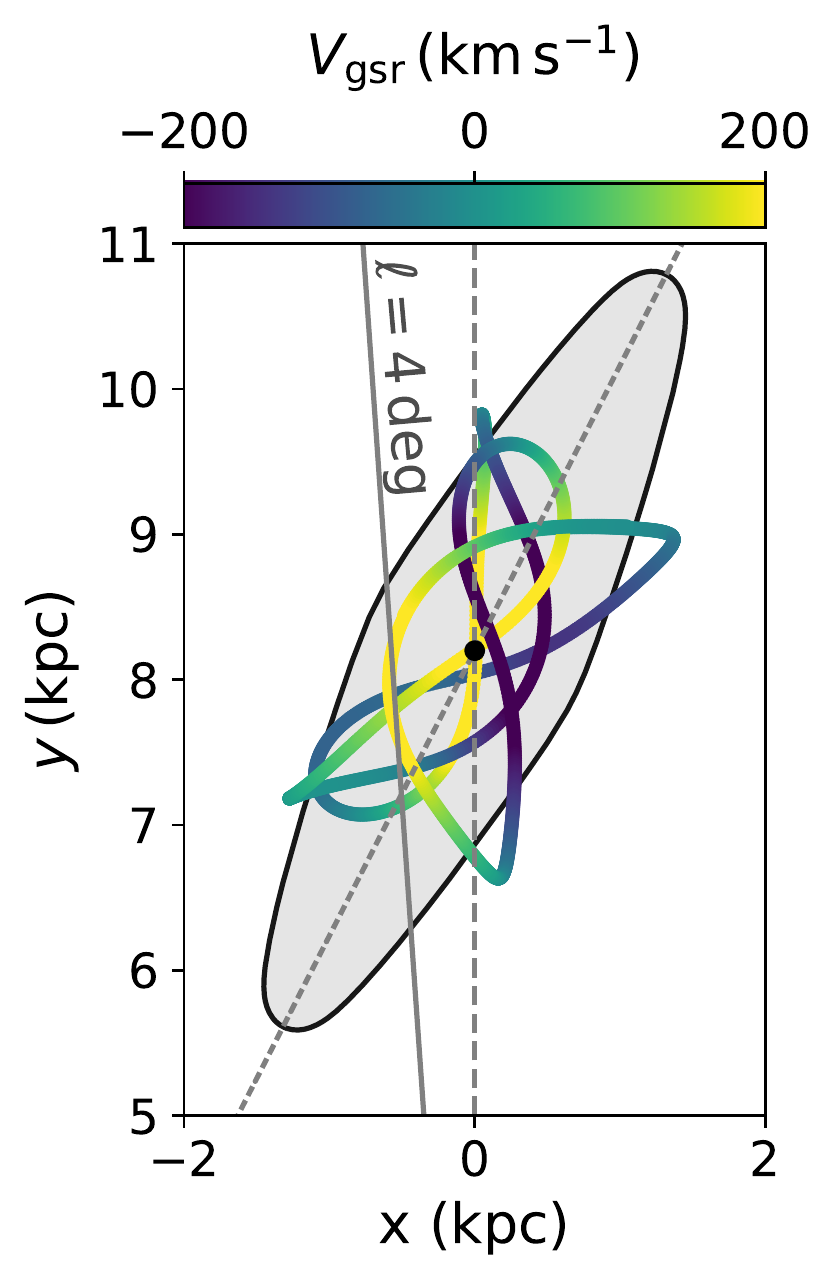}
    \end{minipage}\hfill
    \begin{minipage}{0.8\textwidth}
    \centering
    \vspace{0.27cm}
    \includegraphics[height=6.0cm]{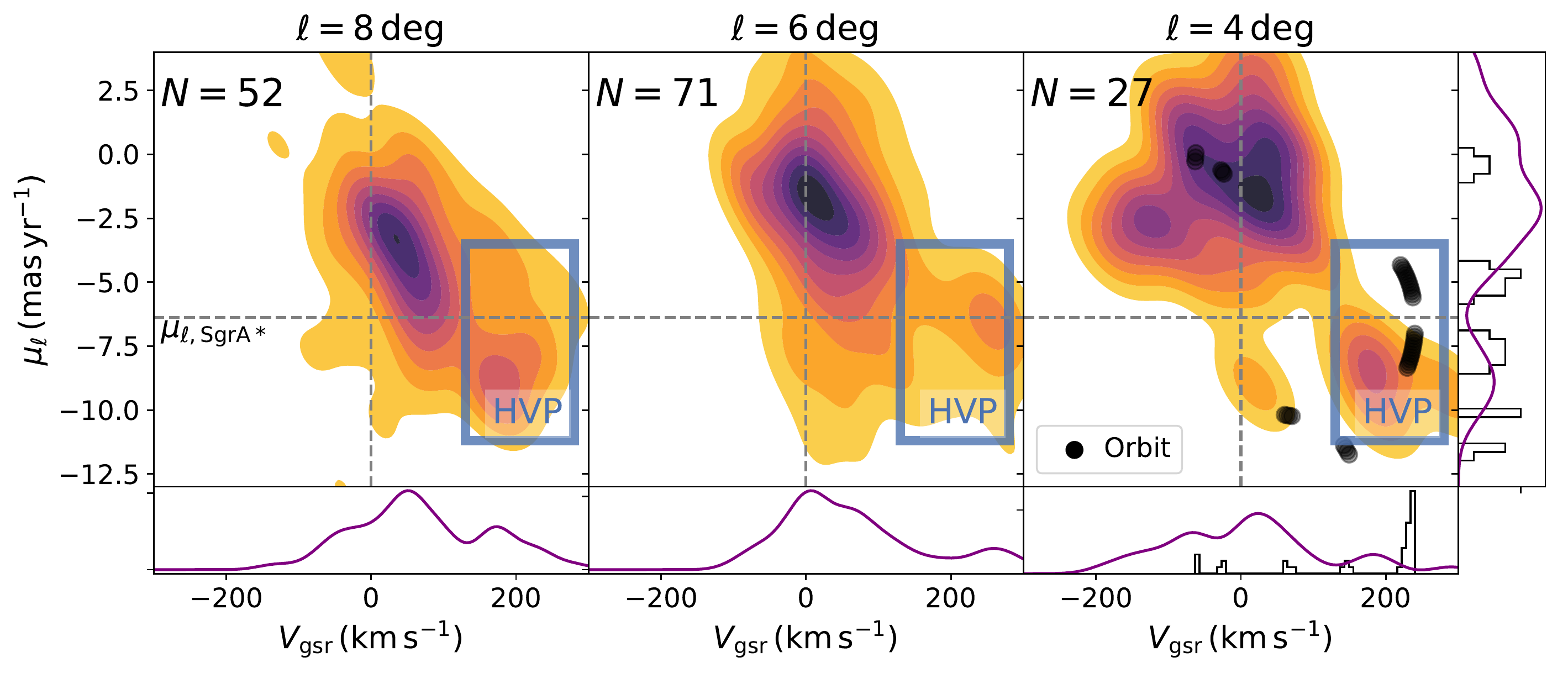}
    \end{minipage}
	\caption{
	A sample propeller orbit showing a characteristic high velocity peak as seen in APOGEE data. This orbit has initial conditions $y=0.544\,\mathrm{kpc}$ and $\dot{x}=235\,\mathrm{km\,s}^{-1}$ ($H_J=-1.207\sqrt{GM/a}$). The left panel shows the orbit coloured by the radial velocity (solid grey line is the $\ell=4\,\mathrm{deg}$ line and the grey shading is an equi-density contour). The right three panels are kernel density estimates of the radial velocity against proper motion space for three APOGEE fields at $\ell=8,6$ and $4\,\mathrm{deg}$ (the number of stars used is given in each panel). Each field shows a high velocity peak. In the $\ell=4\,\mathrm{deg}$ panel we show points from the sample orbit which fall within $0.2\,\mathrm{deg}$ of the line-of-sight, finding this orbit has a high velocity peak that matches in both radial velocity and proper motion.
	}
	\label{fig:velocity_histogram}
\end{figure*}

We take the distance to the Galactic centre as $d=8.2$ kpc and the bar viewing angle as $\alpha = 27^\circ$ \citep{Bl16}. For the velocity of the Sun with respect to the Galactic centre, we use the fact that the longitudinal proper motion of Sgr A* is $-6.379\,\mathrm{mas\,yr}^{-1}$ \citep{Re04}, which, assuming Sgr A* is at rest with respect to the Galaxy, gives the Sun's tangential velocity relative to the Galactic centre as $4.74 \times -6.379 \times 8.2 = -248\,\mathrm{km\,s}^{-1}$. The radial velocity of the Sun towards the Galactic centre is $11\,\mathrm{km\,s}^{-1}$ \citep{Sc10}. We therefore have 
%
\begin{gather}
\vc{v}_\odot = \left(\begin{array}{c} 248 \\ -11 \end{array} \right)\,\mathrm{km\,s}^{-1}.
\end{gather}

A depiction of the geometry of our observation is shown in Fig.~\ref{fig:solar_pov_orbit}. We note that the orbit shown is not the true orbit of any one star as viewed from the Sun, as the bar itself would rotate many times as the star orbits in the bar potential. Instead, the orbit plotted represents an entire family of stars, all on the same orbit, crossing our line of sight simultaneously. 

\subsection{Searching for peaks}

To search for high velocity peaks, we look for orbits that have a large proportion of their radial velocities above a certain threshold. To do this, we integrate orbits with initial conditions $x_0 = \dot{y}_0 = 0$, varying $y_0$ and $H_J$. This will only produce orbits that are symmetrical in the $y$-axis, but this is sufficient for our purposes here. \change{In this way, each orbit corresponds to a point on the left panel of Fig.~\ref{fig:periodic_orbits_for_exp_bar}.} We count the number of times the orbit crosses our observation line with $V_\text{gsr} > 200\,\mathrm{km\,s}^{-1}$, and divide this by the total number of crossings to obtain a ``fast crossing proportion". This is plotted in Fig.~\ref{fig:high_velocity_proportion}. We see there are regions of the orbital space in which high radial velocities are very common. We show a sample orbit in this region, exhibiting the high velocity peaks, in Fig.~\ref{fig:velocity_histogram}. \change{The orbits which can cause these peaks are sired by the periodic orbits D and F (shown in Fig.~\ref{fig:periodic_orbits_for_exp_bar}).}

\change{This is of course only one of a number of orbits that can produce high velocity peaks. The underlying requirement is that the shape of the trajectory close to pericentre must be tangential to the line of sight direction. Other bar models, such as the Ferrers bar studied in \citet{Pa19}, possess high multiplicity periodic orbits that could provide the high velocity peaks. For example, orbits "rm21" and "rm22" in Figure 3 of \citet{Pa19} are superficially similar to our orbit pictured in Fig.~\ref{fig:velocity_histogram}. The proper motion data may allow us to differentiate between a number of plausible possibilities.} 


We close by testing our prediction using proper motion data available from Gaia DR2 \citep{Gaia1,Gaia2} and the VVV proper data \citep[VIRAC][]{Smith2018} tied to an absolute reference frame using Gaia DR2 \citep{Sa19b}.  We take all stars in three APOGEE DR14 fields centred on $b=0$ and $\ell=(8,6,4)\,\mathrm{deg}$ \citep{SDSSDR14}. Following \cite{Zh17}, we remove contaminating foreground stars (in particular disc red clump) by using only $T_\mathrm{eff}<4000\,\mathrm{K}$. We cross-match to the Gaia DR2 catalogue and the VIRAC catalogue (using a cross-match radius of $1\,\mathrm{arcsec}$), and combine the proper motions from the two sources using inverse variance weighting. The resulting (solar-corrected) radial velocity against longitudinal proper motion diagrams are shown in Fig.~\ref{fig:velocity_histogram}. We see all three fields have significant high velocity peaks centred around $\sim200\,\mathrm{km\,s}^{-1}$. The high velocity peak of our candidate orbit produces a peak which coincides well with the peak in the data. 


\section{Conclusions}
\label{section:conclusion}

Analytically tractable bar models are few and far between. This is especially the case for bar models with realistic density profile. We have introduced a novel model for exponential bars, extending the original algorithm of \citet{Lo92}. Our bar model has Gaussian density profiles along the minor and intermediate axes. \change{It has a roughly exponentially falling profile along the major axis, as indicated by the observational date~\citep{El85,Ga07}. It adds extra versatility to the library of analytic bar models in the literature, all of which have density fall-offs that behave roughly like power-laws (e.g., the widely-used Ferrers bars have $\rho \propto (1-R^2/a^2)^n$ along the major axis, so that when $n$ is small the density is more homogeneous than shown by real bars).}

Our bar has four key parameters: $\sigma$ is the variance of the underlying Gaussian density and sets the length scale; $a/\sigma$ is the half-length of the bar; $b/\sigma$ its ``flatness" along the major axis; and $\epsilon$ relates the variances of the intermediate- and minor-axis profiles. We have found analytic formulae for the density, potential and forces arising from this bar, and noted the numerical difficulties that arise. These numerical difficulties do not prevent successful computation of the bar quantities -- our model satisfies $\mathbf{\nabla}\cdot \mathbf{F} = -4\pi G \rho$ to a high degree of accuracy. Furthermore, our model resembles well a Milky Way-like bar obtained from the $N$-body simulation of \citet{Sa19a}. We have therefore succeeded in constructing a bar model that is realistic and customisable, with explicit analytical formulae for the potential and forces. 

The bar has a large parameter space $(\epsilon,a/\sigma,b/\sigma)$ to be explored. We note that our model can give rise to a ``triple well" effective potential, a feature not seen in other bars. Using Poincar\'e sections and by categorising periodic orbits, we have investigated in detail the orbital structure for two different bars. Analysing only these two bars has revealed a variety of novel orbital structure. We have displayed the first bar in which a bifurcating $x_1$ sequence coexists with a stable $x_4$ orbit at all energies. In another bar, the best fit to a simulation of the Milky Way, propeller orbits such as those seen in \citet{Ka05} form the bar backbone. Our analysis suggests that bars with propeller orbits playing a central role are much more common than previously believed. 

We have used this model to investigate the observation of \citet{Ni12}, namely that radial velocity histograms for stars in the Milky Way's bar show distinctive high velocity peaks. We have identified \change{the propeller orbit family as being the likely candidate for such peaks, lending further importance to their role in the orbital structure of exponential bars}. As a test, we have, for the first time, inspected the high velocity peaks distribution in proper motion space, finding a good match with our proposed family. We note that this investigation shows both the value of this model in producing a realistic facsimile of the Milky Way itself, and the ease with which investigations can be conducted using this bar.

A natural extension of the work in this paper is to investigate self-consistency: that is, whether superpositions of different orbital families can reproduce the density distribution. The \citet{Sc79} method was used to show self-consistency of several bar models by \cite{Pf84} and \cite{Ha00}. The key advantage of self-consistency is that it would tell us what proportion of the bar's stellar mass occupies each different orbital family, and explicitly demonstrate the importance of propeller orbits as the major family building the bar. \change{Another fruitful avenue for future exploration is to look for analytic bar models whose vertical profile exhibits the peanut-structure characteristics of buckled disks~\citep[e.g.,][]{Ra91,Lu00} and evident in the Milky Way's deprojected density and stellar populations~\citep{Fr18,Sa19a}.}

Furthermore, while we originally aimed to model those bars with exponentially falling luminosity profiles along the major axis, we have seen that by increasing the parameter $b$, we can obtain very flat luminosity profiles, similar to the flat bar investigated in \citet{Wi17}. In fact, the short flat bar we investigated (with $(a,b) = (1,1)$) gave similar orbital structure to the strong bar in \citet{Wi17}. Our bar has the advantage that it represents only luminous matter in the bar, and not the disc or halo, unlike the bar in \citet{Wi17}, which has a logarithmic potential. This means it can be used in multi-component galaxy models more readily, giving it an extra level of versatility.

\section*{Acknowledgments}
DM would like to thank his co-authors for their limitless patience and expertise in both this paper and the research project that preceded it. The authors thank Leigh Smith for allowing use of his VIRAC v1.1 proper motion catalogue. JLS acknowledges financial support from the Newton Fund, the Leverhulme Trust and the Royal Society.

This work has made use of data from the European Space Agency (ESA) mission
{\it Gaia} (\url{https://www.cosmos.esa.int/gaia}), processed by the {\it Gaia}
Data Processing and Analysis Consortium (DPAC,
\url{https://www.cosmos.esa.int/web/gaia/dpac/consortium}). Funding for the DPAC has been provided by national institutions, in particular the institutions
participating in the {\it Gaia} Multilateral Agreement.

Funding for SDSS-III has been provided by the Alfred P. Sloan Foundation, the Participating Institutions, the National Science Foundation, and the U.S. Department of Energy Office of Science. The SDSS-III web site is http://www.sdss3.org/.

SDSS-III is managed by the Astrophysical Research Consortium for the Participating Institutions of the SDSS-III Collaboration including the University of Arizona, the Brazilian Participation Group, Brookhaven National Laboratory, Carnegie Mellon University, University of Florida, the French Participation Group, the German Participation Group, Harvard University, the Instituto de Astrofisica de Canarias, the Michigan State/Notre Dame/JINA Participation Group, Johns Hopkins University, Lawrence Berkeley National Laboratory, Max Planck Institute for Astrophysics, Max Planck Institute for Extraterrestrial Physics, New Mexico State University, New York University, Ohio State University, Pennsylvania State University, University of Portsmouth, Princeton University, the Spanish Participation Group, University of Tokyo, University of Utah, Vanderbilt University, University of Virginia, University of Washington, and Yale University.




\bsp	
\label{lastpage}
\end{document}